\documentclass{vgtc}                          

\ifpdf
  \pdfoutput=1\relax                   
  \usepackage{graphicx}                
  \DeclareGraphicsExtensions{.pdf,.png,.jpg,.jpeg} 
\else
  \usepackage{graphicx}                
  \DeclareGraphicsExtensions{.eps}     
\fi%

\graphicspath{{figures/}{pictures/}{images/}{./}} 

\usepackage{microtype}                 
\PassOptionsToPackage{warn}{textcomp}  
\usepackage{textcomp}                  
\usepackage{mathptmx}                  
\usepackage{times}                     
\usepackage{cite}                      
\usepackage{tabu}                      
\usepackage{booktabs}                  

\usepackage{setspace}
\usepackage[bookmarks=false]{hyperref}
\usepackage[dvipsnames]{xcolor}
\usepackage[normalem]{ulem}
\definecolor{mred}{rgb}{.80,.12,.30}
\definecolor{MRED}{rgb}{.80,.12,.30}
\definecolor{grey}{rgb}{0.5,0.5,0.5}
\definecolor{lgrey}{rgb}{0.7,0.7,0.7}
\definecolor{purple}{rgb}{.75,0,.85}
\definecolor{pistachio}{rgb}{0.58, 0.77, 0.45}
\definecolor{myorange}{rgb}{0.94, 0.36, 0.13}

\newif\ifnotes
\notesfalse

\newcommand{\smc}[1]{{\small\texttt{#1}}}

\let\origcite\cite
\renewcommand{\cite}[1]{\ifnotes\mbox{\origcite{#1}}\else \origcite{#1}\fi}

\newcommand{\strike}[1]{\ifnotes{\color{mred}{\texorpdfstring{\sout{#1}}{#1}}}\fi}
\newcommand{\strikeg}[1]{\ifnotes{\color{grey}{\texorpdfstring{\sout{#1}}{#1}}}\fi}
\newcommand{\add}[1]{\ifnotes{\leavevmode\color{purple}{#1}}\else{#1}\fi}
\newcommand{\replace}[2]{\ifnotes{\strikeg{#1}\add{#2}}\else{#2}\fi}

\onlineid{0}

\vgtccategory{Research}

\vgtcinsertpkg

\title{Facilitating Exploration with \emph{Interaction Snapshots} under High Latency}

\author{Yifan Wu\thanks{e-mail: yifanwu@berkeley.edu}\\ %
        \scriptsize UC Berkeley %
\and Remco Chang\thanks{e-mail: remco@cs.tufts.edu}\\ %
     \scriptsize Tufts University %
\and Joseph M. Hellerstein\thanks{e-mail: hellerstein@berkeley.edu}\\ %
     \scriptsize UC Berkeley%
\and Eugene Wu\thanks{e-mail: ewu@cs.columbia.edu}\\ %
     \parbox{1.4in}{\scriptsize \centering Columbia University}}

\abstract{
  Latency is, unfortunately, a reality when working with large data sets. Guaranteeing imperceptible latency for interactivity is often prohibitively expensive: the application developer may be forced to migrate data processing engines or deal with complex error bounds on samples, and to limit the application to users with high network bandwidth.
Instead of relying on the backend, we propose a simple UX design---\emph{interaction snapshots}. Responses of requests from the interactions are asynchronously loaded in ``snapshots''.  With interaction snapshots, users can interact concurrently while the \replace{results load asynchronously}{snapshots load}.
Our user study participants found it useful not to have to wait for each result and \add{easily} navigate to prior snapshots\strikeg{easily}.  \strikeg{We found that }For latency up to 5 seconds, participants were able to complete extrema, threshold, and trend identification tasks with little negative impact.

} 


\begin{document}


\firstsection{Introduction}

\maketitle

Current interactive data visualization systems rely on fast response times to provide a good user experience.
This approach simplifies the design of the visualization UI and ensures direct manipulation interfaces that facilitate fluid user data exploration~\cite{liu2014effects}.  However, interactive data visualization is increasingly an integral part of big data analysis.  The scale of the datasets and the required computational power has made it necessary to shift the data processing and storage to remote data\add{bases}\strike{management systems}.  In such a client-server architecture, client interactions are translated into server requests that incur both data processing and network latency.
Ensuring ultra fast response times in the face of all these latencies is often challenging if not impossible.  The interface therefore should have a backup plan should the latency be high---the frontend needs to be resilient to high latencies.

Prior work, such as progressive visualization~\cite{hellerstein1999interactive, fisher2012trust, zgraggen2016progressive, fekete2016progressive, rahman2017ve} and optimistic visualization~\cite{moritz2017trust}, have also utilized interface design to deal with latency.  However, these approaches still rely on considerable backend instrumentation, namely online aggregation~\cite{hellerstein1997online, hellerstein1999interactive, kraska2018northstar} and approximate query processing~\cite{ding2016sample, agarwal2013blinkdb}.
\add{In many settings, designers do not have the opportunity, desire, or resources to make changes to the backend database systems.}

\replace{Critically, all these approaches focus on 
how to address}{Current UX-oriented solutions primarily address} usability challenges stemming from a \emph{single} user request.  They focus on ways to shorten the time between the frontend sending the request to the backend and receiving the response.  For instance, progressive visualization updates a single selected visualization with more accurate results over time.
\replace{When the user want\add{s} to make another interaction while the previous is still being processed, the vast majority of existing designs only allow one interaction to be handled at a time\add{ (with the exception of optimistic visualization, which we discuss in Section~\ref{sec:related})}.
We \add{now} discuss the two most common designs.}{But what happens when the user wishes to make another interaction while the previous is still being processed? We now discuss the two predominant designs.}

One such design is ``blocking'', where users are not allowed to perform a new interaction until the prior ones have rendered.  This design makes the most sense when the latency is negligible, which is often the case when the data is small and fits in memory, such as the case for Vega~\cite{satyanarayan2015reactive} which runs on the browser, and Excel.  The design is easy to implement and puts the least amount of load on the backend.  As a result, even client-server systems like Tableau, which may not always guarantee negligible latency, adopt it.

Another common design is to allow new requests to be made and cancel previous requests.  Allowing the user to interrupt existing requests makes the interface more \emph{responsive{\strikeg{ness}}} and ensures that ``time-consuming operations that block other activity'' can be aborted~\cite{johnson2007gui}.  The interface \strikeg{will }render\add{s} the results of the most recent interaction\strikeg{s} only.

If the previous interaction is \emph{not} cancelled, then more than one pending request will be processed concurrently, which has potential to reduce the overall latency and improve user experience.
However, rendering interaction responses concurrently runs contrary to \textit{direct manipulation}\cite{shneiderman1982future, hutchins1985direct}, a commonly held user-interface design principle.
Direct manipulation requires that ``the object of interest is immediately visible'', which in effect assumes a serial relationship between a user's action and the system's response in a one-to-one fashion.
In contrast, allowing multiple responses to render concurrently behaves in an opposite manner---when a user interacts with a number of visual elements, the system might not respond to these interactions in the sequence the user's actions are performed or to replace the results too quickly. Both of which can be confusing to users by making it difficult to reason about the \emph{correspondence} between interaction and response and to make sense of the responses.

\begin{figure}[b!]
  \centering
  \includegraphics[width=\columnwidth]{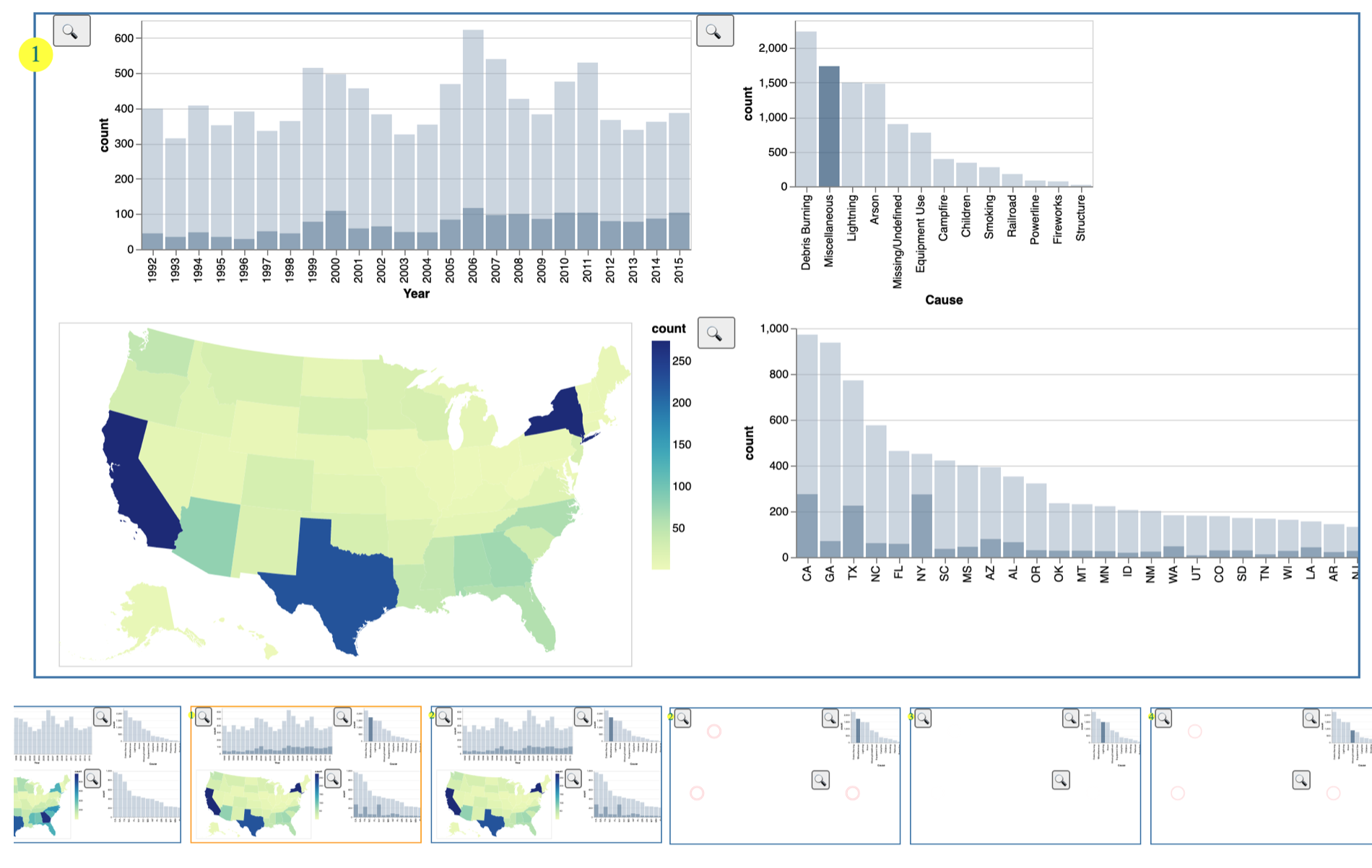}
  \caption{Applying interaction snapshots to a cross-filter visualization. Evaluation of US \replace{wild fire}{wildfire} data.  As users interact, snapshots are created.  Users can perform \replace{``concurrent''}{\emph{concurrent}} interactions where they do not have to wait until the previous results arrive.}
  \label{fig:demo}
\end{figure}

To harness the benefit of concurrent interactions, we must address the design challenge it imposes.  Our approach is to visualize the coordination between asynchronous request and responses explicitly.  We do so by capturing the interaction results in a sequence of \emph{snapshots}.  This way, each new result of an interaction is appended to a history of results. Snapshots provide a stable frame of reference that helps users make sense of uncertain latencies.
Given this easy visual reference, the users could view the snapshots at a later time once the response is received.
Snapshots mediate the asynchronous results through history and create a direct manipulation experience for multiple concurrent interactions.

Consider a cross-filter application, as shown in Fig.\ref{fig:demo}.  After a user makes an interaction, a snapshot is created and appended beneath the visualization dashboard.  The snapshot is a scaled-down display of the current visualization which continues to load while other snapshots are appended.
The snapshot gives a visual indicator (e.g., spinner) of whether the data is still being processed.  When the user sees the visual indication that the processing is complete, they can click on the corresponding snapshot, which loads the processed visualization into the main view for analysis.
Users can also navigate through the snapshots quickly with left and right arrows, which ``animates'' through the selections.

We evaluated the efficacy of interaction snapshots on dashboards with 6 participants. Traces of participant behavior demonstrate they can make effective use of concurrent interactions and navigate to different snapshots.  Qualitative feedback reveals that participants find the interaction snapshot design helpful in the face of latency.



\section{Related Work}
\label{sec:related}

\noindent{\textbf{Dealing with Interactive Latency:}} Prior work addressing the issue of latency in interactive visualizations can be divided along two axes: whether the solution is provided by the backend or the frontend, and whether the query is evaluated on whole or samples of the data.

Backend techniques to reduce latency for queries over all of the data include GPU-based compute~\cite{liu2013immens, mapd, graphistry}, custom indices~\cite{lins2013nanocubes, tao2019kyrix}, and prefetching~\cite{moritz2019falcon, battle2016}.
Backend techniques for a probabilistic query trade some amount of uncertainty for the reduction in processing time.  These include approximate query processing~\cite{agarwal2013blinkdb, ding2016sample} and online aggregation engines~\cite{hellerstein1997online, hellerstein1999interactive} from which \emph{progressive visualization}~\cite{hellerstein1997online, fisher2012trust, zgraggen2016progressive, fekete2016progressive, rahman2017ve}, which we discuss in the next paragraph, is based on.
While innovations in these backend techniques improve computation capabilities, they do not eliminate the existence of latency in the real world.  Users could be dealing with legacy systems, large amounts of data, or, sometimes, slow network connection.

Compared to the pure backend efforts, our work is more closely related to progressive visualization, which \strikeg{develops around the idea of }\replace{streaming}{streams} partial results (containing error bounds) to users.
\add{An augmentation to the streaming design}
\strikeg{Another related idea} is  Moritz et al.'s \emph{optimistic visualization}, which allows the user to first interact with an approximate query engine, and lets users mark an interaction as ``remembered'' for a full, non-approximate, evaluation to check later~\cite{moritz2017trust}.

We take inspiration from these work's approach of leveraging design to adapt to the realities of ``big data''. 
\add{In particular, optimistic visualization allows for users to optimize \emph{across} interactions, from which our design builds on.}
However, both progressive and optimistic visualization rely on approximate query processing.
Since on-the-fly sampling cannot cover every small subset of data, many approximate query techniques also involve precomputing samples, sketches, or other summary structures~\cite{cormode2012synopses}. The preprocessing steps require time, computation, and storage for each precomputed result.
\add{The additional effort may be worthwhile for developers who could afford backend changes, e.g., adapting advanced engines like \emph{Sample+Seek}.  For those who would rather make changes just to the frontend, interaction snapshots may be a more preferable trade-off---developers just need to add the new UX technique and potentially limit their interaction designs to avoid ones that would trigger a large amount of interactions, such as continuous brushing.}
\strikeg{Interaction snapshots, as a simple UX technique, does not depend on these more involved backend efforts.}

\medskip
\noindent{\textbf{Interaction History:}}
Much prior HCI work has used interaction histories to facilitate user actions.  
Work in the CSCW community used trails of cursor positions to give temporal context to the actions of remote participants~\cite{savery2011s}.
In the visualization community, Heer et al. model history as a sequence of movements through a graph of application states, presented in thumbnails in \emph{Graphical histories}~\cite{heer2008graphical}.
Feng et al. externalized interaction history by showing the ``footprints" of interactions\cite{feng2017hindsight}. Optimistic visualization, as mentioned earlier, also makes use of history (the ``remember'' feature) to help users verify approximate results~\cite{moritz2017trust}.

These prior work inform our design.  Feng et al. observed that a direct encoding of interaction history supports visual recognition of previous interactions. The visual history does not require users to recall the past, which can be mentally taxing.  The observation inspired us to visualize interaction history to reduce the cognitive challenges to asynchronous interactions.  Graphical histories gave us inspirations for how to design the snapshot for dashboards, and optimistic visualization gives initial support that users do revisit interaction history when exploring data.

\section{Design Iterations}

\begin{figure}[tb]
   \centering
   \includegraphics[width=0.8\columnwidth]{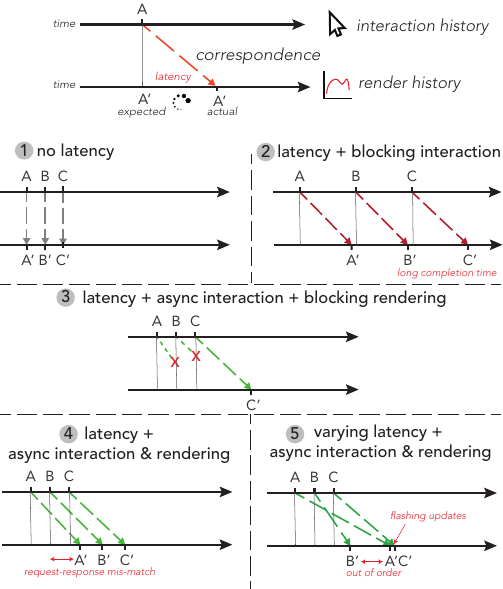}
   \caption{A sequence of interaction requests and responses under different conditions visualized on a horizontal time axis.  Colored arrows represent request/response pairs over time. Light vertical lines highlight request times. Case (1) is the ideal no-latency scenario commonly assumed by visualization designers---everything works as expected.  (2) With latency, the user waits for each response to load before interacting.  (3) With latency, the user interacts without waiting, and in-flight responses are not rendered.  (4) With latency, the user interacts without waiting, and all responses are rendered.  (5) With latency, the user interacts without waiting and may see responses in a different order than requests were issued.}
   \label{fig:events}
\end{figure}

To design with latency, we first analyze different ways interactions are handled in the presence of latency.
The top diagram in \replace{Figure}{Fig.}~\ref{fig:events} depicts a time-ordered model, where time increases from left to right. User inputs are depicted along the top line (interaction history), and the responses are rendered along the bottom line (render history). A dashed arrow between the interaction and render history corresponds to the time to respond to the request.

\begin{figure}[b]
   \centering
   \includegraphics[width=\columnwidth]{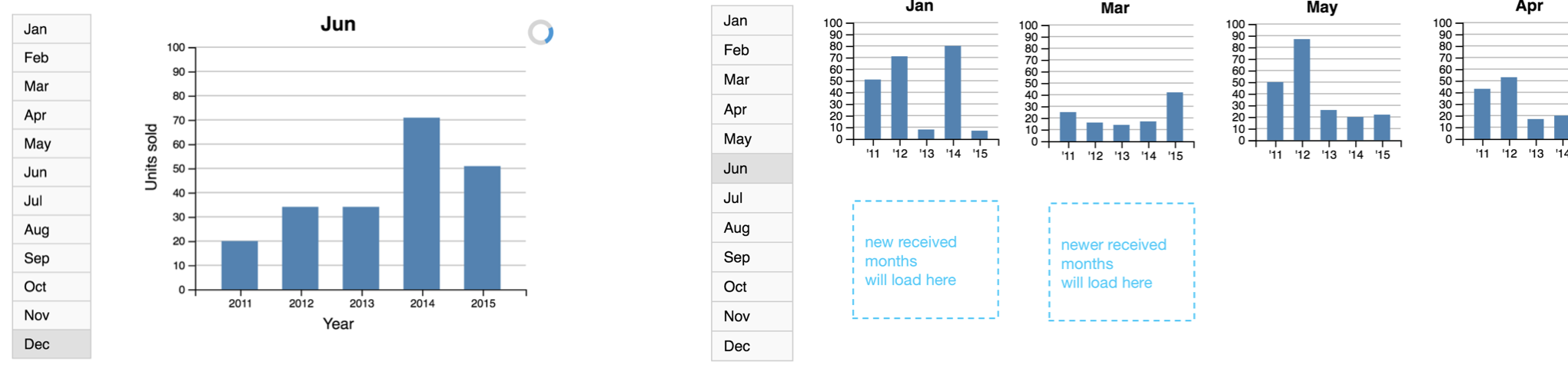}
   \caption{Pilot experiment: on the left is the basic design where interaction results update in place, on the right is a design that displays snapshots of interaction results as small multiples.}
   \label{fig:pilot}
\end{figure}

\begin{figure}[tb]
   \centering
     \includegraphics[width=0.7\columnwidth]{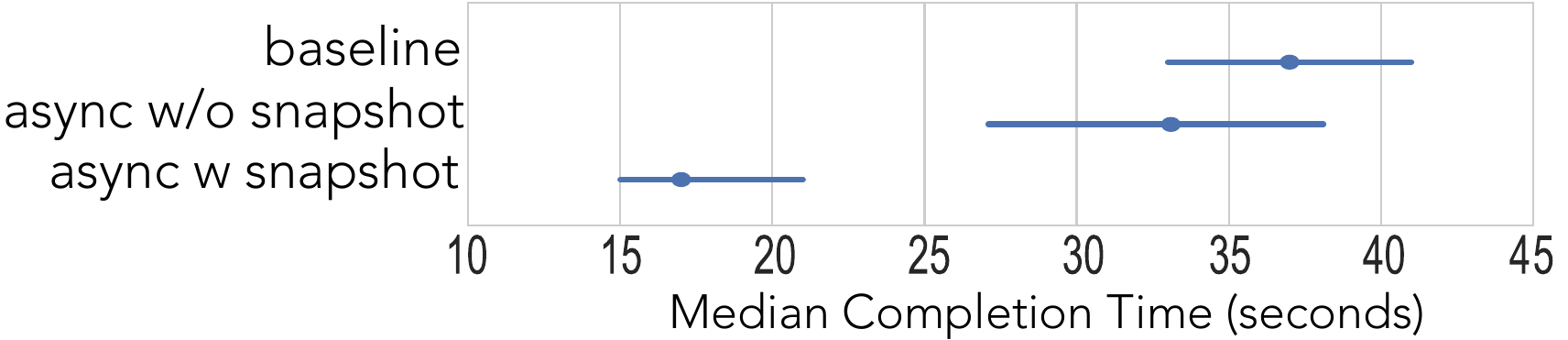}
   \caption{Comparison of median users task completion times, with the \emph{interaction snapshots} condition being much faster than the others.}
   \label{fig:pilot_time}
\end{figure}

\replace{Figure}{Fig.}~\ref{fig:events}(1) shows the ideal case where requests respond instantaneously. However, \replace{assuming}{when} there is latency\strike{ in the system}, a number of possible scenarios can occur: \strikeg{For example, }(2) shows the blocking case where the user is not allowed to submit a new request until the prior one\replace{ is rendered}{completes}; (3) shows the non-blocking case where users freely interact with the visualization and new requests supersede and cancel previous requests; (4) \replace{illustrates}{shows} the concurrent case where neither the input nor the output is blocked. The benefit of this approach is that the total time is shorter, but the downside is that the interface could be difficult to interpret, especially when the amount of latency varies\strikeg{, as shown in } (5).

\begin{figure}[tb]
   \centering
     \includegraphics[width=0.7\columnwidth]{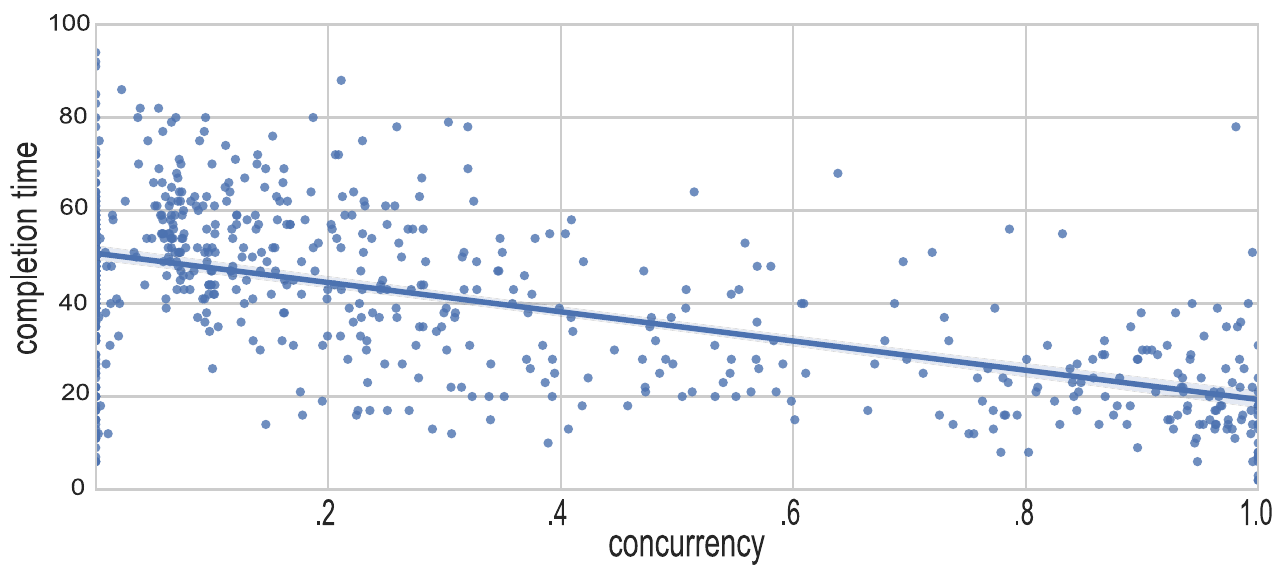}
   \caption{Completion time correlated with level of concurrency. A negative correlation suggests that concurrent interactions may help alleviate the effect of latency.}
   \label{fig:pilot_concurrency}
\end{figure}

\begin{figure}[tb]
   \centering
   \includegraphics[width=0.75\columnwidth]{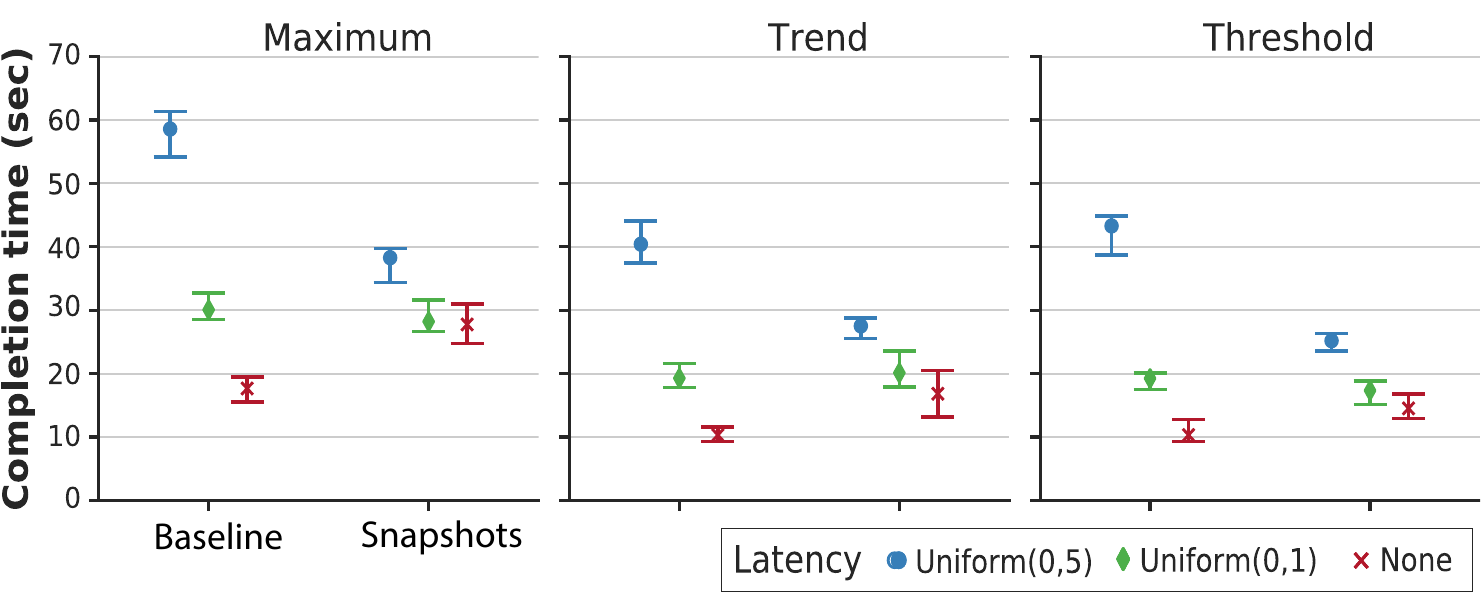}
   \caption{Each chart of the plot visualizes median task completion time with 95\% CI (y-axis).}
   \label{fig:exp2_all}
\end{figure}

To address the dis-coordination between interaction request and response, as seen in \replace{Figure}{Fig.}~\ref{fig:events}(4,5), we hypothesize that displaying past interactions in snapshots could serve as a stabilizing visual anchor.  
A simple mechanism is to encode the step by which the interaction was made using a\strikeg{n additional} visual encoding channel.  The right of Fig.~\ref{fig:pilot} shows an example where \strikeg{snapshots of month }selections are rendered \add{in snapshots}.

We conducted a pilot study to verify the hypothesis.  In the pilot, participants used a simple visualization shown in Fig.~\ref{fig:pilot}: a bar chart that displays sales data for a company across years, split into facets of the months.
There are three conditions: baseline, treatment 1 and treatment 2. The first two uses the interfaces on the left of Fig.~\ref{fig:pilot}, and the last one the right of Fig.\ref{fig:pilot}.
The baseline is the blocking interaction, as illustrated in Fig.\ref{fig:events}(2). Treatment 1 has the same UI as baseline, but asynchronously renders the results, as in Fig.\ref{fig:events}(4,5).
Treatment condition 2 shows interaction snapshots.

Participants were asked to identify if any of the months crossed the sales threshold of $80$ units sold. 
We measured the accuracy of the response and the total time to complete a task in seconds \add{(the time between when the participant is allowed to start interacting and when the they submits an answer)}.  We also logged all events on the UI, such as \strikeg{hover }interactions, responses received, and response rendered.


We recruited participants online through Amazon Mechanical Turk (17 participants for baseline, and 30 for the two treatments, 58\% with bachelors degree or higher, and 46\% female, ranging from 23 to 67 years of age).  \add{Participants were compensated \$0.30 per task, with a \$3-5 completion bonus, compliant with Californian minimum wage.} Participants were randomly sorted into either the baseline or treatment group.  They were shown instructions about the task and trained to complete two sample \replace{assignments before completing actual assignments}{tasks beforehand}.

The differences were in task completion time, shown in Fig.~\ref{fig:pilot_time}.  We report the unsigned Wilcoxon Rank-Sum test: baseline median=37 sec (N = 31), treatment (condition 1) without snapshots median=33 sec (N=52), Z=0.63, $p<0.5$, and treatment (condition 2) with snapshots median=17 sec (N=54), Z=3.22, $p<0.002$ where N denotes the count of the group.  There were no significant \strikeg{accuracy }differences \add{in accuracy} between the three conditions.  \add{We can see that participants were able to complete the tasks much faster with the snapshots design.  To understand why this was the case, we visualized the concurrency---the percentage of task completion time where there was more than one concurrent request. Fig.~\ref{fig:pilot_concurrency} show that participants \strikeg{could complete tasks faster }with higher concurrency\add{ tend to complete tasks faster}, which is made possible because of asynchrony and encouraged by the snapshot design.}

To ensure that the result also generalizes to other tasks, we conducted a second pilot study to include two more tasks: identifying the month with a maximum value, and the month with a certain trend. These two tasks represent the ``find maximum/extremum'' and ``characterize distribution'' tasks in Amar et al.'s analytic activity taxonomy \cite{amar2005low}. Both of these tasks are known to be more challenging than the threshold task in the first pilot study, which falls under ``retrieve value'' in the taxonomy.  Each participant completed the tasks with either no snapshots (baseline) or with snapshots designs (treatment).  We also have two latency conditions: one is uniformly sampled from 0 to 1 second (short), and the other is  uniformly sampled between 0 to 5 seconds (long).  Each participant completes three conditions (none, short, and long).  For each group, we recruited 50 Mechanical Turk participants.  Again, we found that there were no significant differences in task accuracy.  Task completion time, however, was very different, as shown in Fig.~\ref{fig:exp2_all}.  We see that participants complete all tasks significantly faster in the with-snapshots condition when there was \add{long} latency. 


More qualitatively, participants commented that completing the tasks with latency with no snapshots is ``painful'', ``frustrating'', ``tedious'', and ``awful''.  Some explained that responses were hard to remember---\emph{``I had a hard time remembering what I'd just seen a second ago.''}.
In contrast, participants commented on the ease of use when snapshots are present---\emph{``The ability to load several months at once definitely offsets any loading latency -- difficulty was roughly the same as one month with no latency. One month with latency was a bit painful.''}.  Interestingly, the perceived speed of loading seemed to have changed as well---\emph{``Some of the tasks loaded really slow, single month got irritating waiting.  Most of the multiple tasks loaded fairly quickly.''} While the perception of latency is not the focus of this study, \replace{such participant}{the} feedback suggests the benefits of the use of \replace{``interaction snapshots''}{interaction snapshots} beyond improving \strikeg{the participants' }task-completion times~\cite{harrison2007rethinking}. 

\begin{figure*}[tb]
  \centering
  \includegraphics[width=\textwidth]{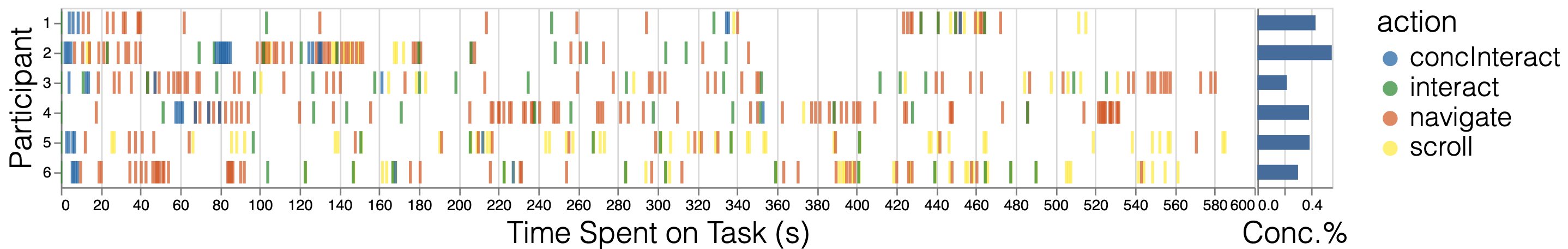}
  \caption{On the left is a visualization participants' interaction traces while exploring  US \replace{wild fire}{wildfire} data. The traces visualized are interactions---which contains both concurrent interactions (\smc{concInteract}), and non-concurrent interactions (\smc{interact}).  On the right is the percent of concurrent interactions of all interactions made.}
  \label{fig:eval_trace}
\end{figure*}

\replace{Snapshots}{The interaction snapshots design} was not free from fault. One participant commented that ``It took a few tries to get used to how it worked''.  We also see evidence of this in \add{Fig.~\ref{fig:exp2_all}---under} \strikeg{the results for }the no-latency condition, \strikeg{where }participants took on average \strikeg{slightly }longer to complete \strikeg{the maximum }task\add{s} when using the new design compared to baseline.
\section{Dashboard Snapshots: Design and Evaluation}

The pilots evaluated interaction snapshots' effectiveness on a single visualization for fixed tasks.  We now evaluate the technique for a more complex setting---dashboard---with open ended exploration.

One key design challenge with snapshots for dashboard is \emph{space}.  Replicating the interaction results as small multiples is not feasible for dashboards.  We address this constraint by creating a separate representation of the snapshot---a smaller ``thumbnail'' view, much like Graphical Histories~\cite{heer2008graphical} and Pangloss~\cite{moritz2017trust}, which can be clicked on and expanded. 
Figure~\ref{fig:demo} is an example application of the technique on a cross-filter visualization of wildfires in the US~\cite{fires}.
\add{The interaction details and code are included in supplementary materials.}

\subsection{Methods}

Rather than assigning participants specific tasks (as in previous pilots), we \strikeg{plan to }observe how participants explore data and \replace{use more}{report} qualitative metrics.  First is whether and how much the \replace{analysts}{participants} make use of concurrent multiple interactions.  A higher usage would suggest that the snapshots design is able to facilitate concurrent interactions, and that it is useful to the participants.
Second is how often snapshots are revisited\replace{.  Since traditional dashboards reduce to the most recent snapshot, }{---}the more participants engage in older snapshots, the more they are leveraging interaction snapshots' unique capabilities.
Third is direct feedback.

We conducted a first-use study with 6 \strikeg{representative }users, all college students who have taken data science courses, with a self-reported ``somewhat experienced'' with visual data analysis on a 5-point Likert scale $(\mu=3.0, \sigma=0.89)$.
Due to the COVID-19 ``shelter-in-place'' order, we conducted the studies over video.
We began each study with a 5-minute tutorial of the interaction snapshot enabled dashboard on mass mobilization protest data~\cite{DVNHTTWYL2016}, and then asked participants to \replace{analysis}{analyze} US \replace{wild fires}{wildfires} using a similar dashboard (Fig.~\ref{fig:demo}).
Participants \replace{are}{were} \replace{promoted}{prompted} with a specific question (``identify the states with the most \strikeg{number of }wildfires in the years 2000 to 2004'') followed by free-form exploration for about 8 minutes. \replace{At the end}{Then}, participants verbally summarize\add{d} their findings.
All interaction latency \replace{is}{was} set to between 5 to 7 seconds.
At the conclusion of the study, we administered an exit survey to measure the effectiveness of the interface and to debrief participants about their experiences.
The \strikeg{whole }session took 20-25 min and participants were compensated \strikeg{with }\$15 in Amazon gift cards.

\subsection{Quantitative Results}

We instrumented the interface to log all user interactions, including interactive selections of elements in the charts, navigating to prior interaction states, and scrolling through the snapshots.
To analyze this data, we visualized the traces in Fig.~\ref{fig:eval_trace}.  For the interactions that happened \emph{before} the previous interaction was loaded (since there is a 5 to 7 second delay), we mark them as \emph{concurrent interactions} (\textcolor[HTML]{0868ac}{\smc{concInteract}}, as labeled in the Chart)
the rest of the interactions as \textcolor[HTML]{239023}{\smc{interact}}.  We also provide a distribution of the percent of concurrent interactions out of all interactions in the bar chart to the right, with ($\mu=0.38, \sigma=0.10$).  Participants interacted with the interface on average 20 times $(\mu=20.33, \sigma=5.99)$ during the session, and navigated twice that \add{on }average $(\mu=43.17, \sigma=17.49)$.


On 5-point Likert scales, participants positively rated the interface overall $(\mu=4.33, \sigma=0.47)$, as well as the history feature $(\mu=4.17, \sigma=0.69)$.  In terms of how frustrating the delay was, participants rated it as only a little $(\mu=2.0, \sigma=0.58)$.


\subsection{Qualitative Results}

We observed participants quickly grasped how to make use of concurrent interactions through the snapshots.
One common pattern was to make multiple interactions concurrently when the participants had a question in mind.  P3 mentioned that it was ``\textit{nice to have [the interaction result] pre-loaded}'', and that reminded them of ``\textit{opening search results in multiple tabs [in the background] when browsing web pages}.''  P6 also indirectly commented that ``\textit{Tabs [i.e., snapshots] made the wait less painful/annoying.}''

However concurrent interactions are not always used. When participants had a specific question or targets in mind, such as ``I want to compare how different causes of fires differ geographically'', they knew exactly what interactions they would need and opted for concurrent interactions.  When the participants did not have such a question, they relied on the results to their immediate selections to generate ideas for further interactions. Hence they waited for the response to load instead of making other interactions while waiting. 
Interestingly, the snapshots still proved useful while they \emph{waited}.  P4 mentioned that when they were waiting for the result to load, they would ``\textit{look at the history to remember what I was doing earlier to keep track of what I'm looking for}''.  P1 also used the delay to positive effect, saying that ``\textit{the delay gives my brain a time to reflect on what I'm expecting and what to do. I think it's actually better [than instantaneous response]}.''  None of the participants found the delay to be more than a little frustrating.

All of the participants \replace{used navigation}{browsed} through history when summarizing their findings\replace{.  For the final task of summarizing findings, they browsed through history}{ (the final task)}, recalling relevant insights they made prior.  This can be seen in the dense patches of \textcolor[HTML]{d95f0e}{\smc{navigate}} ticks towards the end of the sessions in Fig.~\ref{fig:eval_trace}.
P6 mentioned that ``\textit{History tool allowed me to go back to my previous thoughts easily, and made it easy to reference observations.}''

\strikeg{However, }Participants also desired more features.  P4 mentioned that the snapshots quickly became visually cluttered and difficult to navigate, which detracted from the positive aspects of history.
P3 suggested ways to introduce more guides for the snapshots, such as adding text to describe \strikeg{what }the interaction \strikeg{was}, or a way to either arrange or color encode the snapshots by the chart interacted with.
\strikeg{Both }P3 and P4 asked about the ability to remove snapshots that they found irrelevant.

\section{The Interaction Snapshot Design Process}

Having presented \replace{ways to create snapshots of interactions by example}{examples of interaction snapshots}, we now conclude with a generalized design process. Interaction snapshots  require three elements: (1) the user's past interactions, (2) the effect of each interaction, and (3) the temporal ordering of the interactions.
\replace{These three components }{Together, they} show the correspondence between the user's interaction \add{requests} and the response of the system over time.
There are different ways to satisfy these requirements. 
For the \strikeg{pilot }bar chart, we used \replace{Bertin's ``position'' visual}{the position encoding} channel\add{, which can be applied to other single visualizations.}\strikeg{, and }
For the dashboard, we used \replace{thumbnails}{snapshots}\add{, which can be applied to other multiple coordinated visualizations}. 
\strikeg{These two designs in fact generalize to most common interactions---either single chart interactive visualizations or multiple coordinated visualizations.}

Through many pilots and design iterations, we explored aspects of the design space for multiple concurrent interactions.  We found a positive answer to our initial question of whether frontend design alone could offer some alleviation to the pain of latency.
We have also opened up new questions.  One idea is to more systematically study how snapshots change user behavior in terms of rates of observations, drawing generalization, and forming hypothesis\add{, following prior work}~\cite{liu2014effects}\replace{.  We could also}{ and} compare interaction snapshots and progressive/optimistic visualizations.
Another idea is to further develop controls around the snapshots\replace{---participants voiced interest\strikeg{s} in}{such as} editing and organizing\strikeg{ the snapshots}.  \add{We could also augment the linear history with when users ``branched'' off into a different interaction to capture richer context---the snapshots could double as an interaction provenance graph.}
More broadly, \emph{interaction snapshots} present\strikeg{s} opportunities to bring features common in literate computing to interactive visualizations\strikeg{that are usually transient}.


\acknowledgments{Thanks to Larry Xu for early contributions, Aditya Parameswaran and the anonymous reviewers for their constructive comments. This material is based upon work supported by: NSF 1564049 1845638, 1564351, CCF-1730628, OAC-1940175, OAC-1939945, IIS-1452977, DGE-1855886, DARPA D3M (FA8750-17-2-0107), DOE DE-SC0016934, and gifts from Alibaba, Amazon Web Services, Ant Financial, CapitalOne, Ericsson, Facebook, Futurewei, Google, Intel, Microsoft, Nvidia, Scotiabank, Splunk and VMware.}

\bibliographystyle{abbrv-doi}

\bibliography{ref}
\end{document}